\documentclass[epsfig,showpacs]{revtex4}
\usepackage{epsfig,color}

\begin{document}
\title{Recent astrophysical and accelerator based results on the
Hadronic Equation of State}

\author{Ch. Hartnack${}^{1}$, H. Oeschler${}^{2}$ and J\"org Aichelin${}^{1}$
\footnote{invited speaker}}

\address{${}^{1}$SUBATECH,
Laboratoire de Physique Subatomique et des Technologies Associ\'ees \\
University of Nantes - IN2P3/CNRS - Ecole des Mines de Nantes \\
4 rue Alfred Kastler, F-44072 Nantes Cedex 03, France}

\address{${}^{2}$Institut f\"ur Kernphysik,
Darmstadt University of Technology, 64289 Darmstadt, Germany}

\begin{abstract}
In astrophysics as well as in hadron physics progress has recently been made 
on the determination of the hadronic equation of state (EOS) of compressed 
matter. The results are contradictory, however.
Simulations of heavy ion reactions are now sufficiently robust to predict
the stiffness of the (EOS) from (i) the energy
dependence of the ratio of $K^+$ from Au+Au and C+C collisions and
(ii) the centrality dependence of the $K^+$ multiplicities.
The data are best described with a compressibility coefficient
at normal nuclear matter density $\kappa$
around 200 MeV, a value which is usually called ``soft''
The recent observation of a neutron star with a mass of twice the solar mass
is only compatible with theoretical predictions if the EOS is stiff.
We review the present situation.
\vspace{.6cm}
\end{abstract}

\maketitle

How much energy is needed to compress nuclear matter? The answer to this
question, the determination of $E/A(\rho,T)$, the
energy/nucleon in nuclear matter in thermal equilibrium as a
function of the density $\rho$ and the temperature $T$,
has been considered since many years as one of the most important
challenges in nuclear physics.  This quest
has been dubbed ``search for the nuclear equation of state (EoS)''.

Only at equilibrium density, $\rho_0$, the energy per nucleon
$E/A(\rho=\rho_0,T=0)= -16$ MeV is known by extrapolating the
Weizs\"acker mass formula to infinite matter.
Standard ab initio many body calculations do not allow for a determination
of  $E/A(\rho,T)$ at energies well above the saturation density
because the low density many body expansion schema ( Br\"uckner G- matrix) 
breaks down and therefore the number of contributing terms is exploding.
Therefore in nuclear reaction physics another strategy has been developed. Theory has identified
experimental observables in nuclear reaction physics or in astrophysics
which are sensitive to $E/A(\rho,T)$. Unfortunately these observables depend as
well on other quantities which are either unknown or little known (like cross
sections with resonance in the entrance channel) or difficult to asses
theoretically (like the resonance lifetimes in hot and dense matter).
It was hoped that comparing many observables for different systems
and different energies with the theoretical predictions these unknown or
little known quantities can be determined experimentally and that finally the
dependence of the observables on $E/A(\rho,T)$ can be isolated.

\begin{figure}
\vspace*{-.5cm} 
\epsfig{file=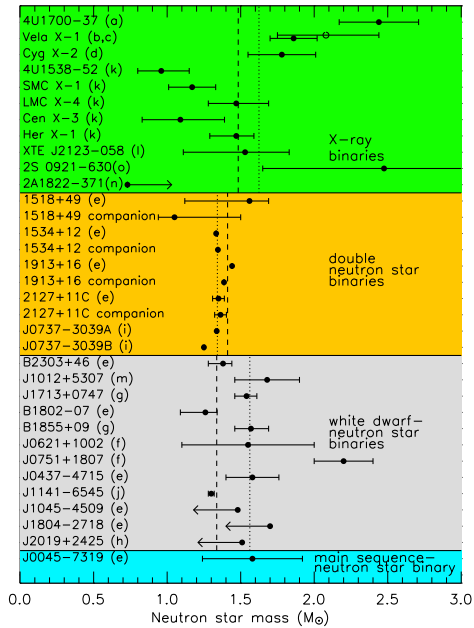,width=7cm} \caption{Measured and estimated
masses of neutron stars in radio binary pulsars and in x-ray
accreting binaries.  Error bars are $1\sigma$.
Vertical dotted lines show average masses of each group (1.62
M$_\odot$, 1.34 M$_\odot$ and 1.56 M$_\odot$); dashed vertical lines
indicate inverse error weighted average masses (1.48 M$_\odot$, 1.41
M$_\odot$ and 1.34 M$_\odot$). The figure is taken from ref \cite{lat1}}
\label{mas}
\end{figure}

\begin{figure}
\vspace*{-.5cm} 
\epsfig{file=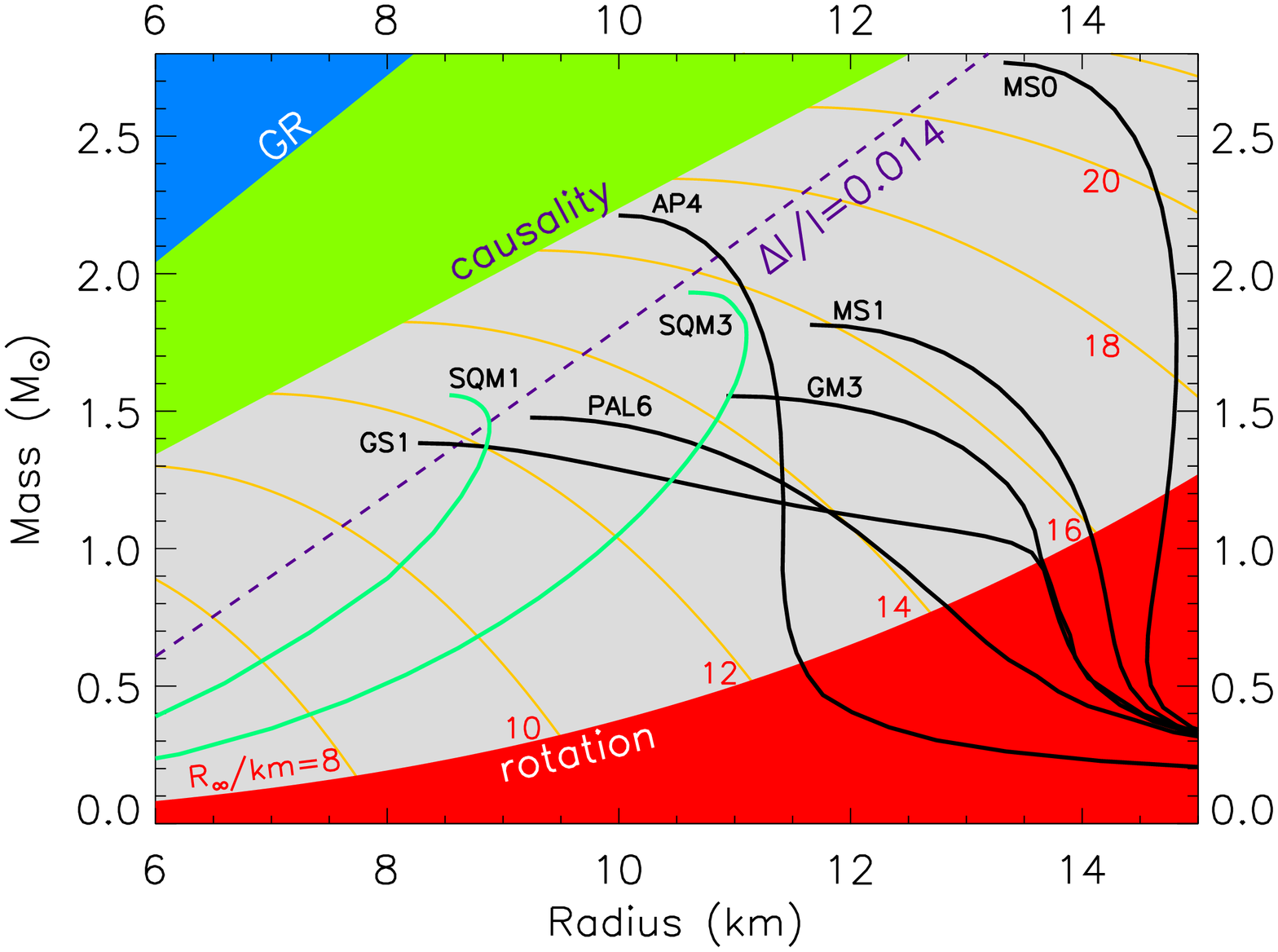,width=12cm,angle=90} \caption{Mass-radius
diagram for neutron stars.  Black (green) curves are for normal
matter (SQM) EoS [for definitions of the labels, see
\cite{lat1}].  Regions excluded by general relativity (GR),
causality and rotation constraints are indicated. Contours of
radiation radii $R_\infty$ are given by the orange curves. The figure
is from  \cite{lat1}.} \label{mr}
\end{figure}

In astrophysics the nuclear EoS plays an important role
in binary mergers involving black holes and neutron stars \cite{merger},
in double pulsars \cite{pod}, in the mass-radius relation of neutron stars
\cite{lat,lat1} and in supernovae explosions \cite{jan}. For a recent review on
these topics we refer to \cite{weber}. Unfortunately, as in nuclear reaction
physics, there are always other little known processes or properties which
have to be understood before the nuclear EoS dependence can be
isolated. We discuss here as example of the mass-radius relation of neutron
stars. Fig. \ref{mas} shows the neutron star masses in units of the solar mass
for different types of binaries. These masses are concentrated at around 1-1.5 solar
masses.
Fig. \ref{mr} shows a theoretical prediction of the mass-radius relation
for neutron stars using different EoS. Since the nature of the
interior of neutron stars is not known (in contradiction to what the name
suggests) one may suppose that it consists
of hadrons or of quarks. But even if it consists of hadrons there are
speculations that there is a $K^-$ or a $\pi^-$ condensate or that there are
hyperons in equilibrium with nuclear resonances. The same is true if the
interior consists of quarks. Little known color-flavor locked quark phases
may modify the EoS at densities which are reached in the interior of the
neutron star.
For a detailed discussion of all these phenomena we refer to ref. \cite{weber}.
We see that the observed masses of neutron stars are compatible with almost
all quark or hadron based EoS as long as the radius is unknown.
Radii, however, are very difficult to measure. Because similar problems appear
also for other observables, up to recently
the astrophysical observations of neutron stars did not help much to narrow down
the uncertainty on the nuclear EoS.

This situation has changed dramatically in
the last year with the observation of a neutron star with a mass of
two solar masses \cite{star}. If this observation is finally confirmed the
mass/radius prediction of fig.\ref{mr} excludes that the interior of a neutron
star is made by quarks \cite {lat1}, even a soft nuclear EoS,
which will be defined below, will be excluded. This is confirmed by the
calculation of Maieron \cite{mai} which uses a MIT bag model or a color dielectric
models EoS to describe the quark phase. Baldo \cite{bal} argue
that this conclusion may be premature because it depends too much on the
equation of state of the quark phase. If one replaces the MIT bag model
equation of state by that of the Nambu - Jona-Lasinio (NJL)  Lagrangian under
certain conditions (no color conducting phase) larger masses may be obtained.
The standard NJL Lagrangian lacks, however, repulsion and
in view of the momentum cut-off, necessary to regularize the loop integrals,
and the coupling constants in the diquark sector, which are not uniquely
determined by the Fierz transformation, quantitative prediction at high
quark densities are difficult in this approach even if qualitative agreement
with pQCD calculation can be found \cite{gas}.

Simulations of heavy ion reactions have shown that there are three possible
observables which are sensitive to $E/A(\rho,T)$ at densities larger 
than $\rho_0$:
(i) the strength distribution of giant isoscalar monopole resonances \cite{you,mon},
(ii) the in-plane sidewards flow of nucleons in semi-central heavy
ion reactions at energies between 100 $A$ MeV and 400 $A$ MeV
\cite{sto} and 
(iii) the production of $K^+$ mesons in heavy ion
reactions at energies around 1 $A$ GeV \cite{aik}.
For the present status of these approaches we refer to \cite{prl}.
 
Monopole resonances test the nuclear EoS at densities only 
slightly larger than the normal nuclear matter density. Therefore they are of
little help if one compares the EoS determined from astrophysics with that
extracted from nuclear reaction physics. For the in-plane flow the conclusions
are not conclusive yet. 
This is due to the difficulties to determine
the EoS in heavy ion collisions. An EoS is defined in a thermally equilibrated
system but in heavy ion collisions equilibrium is not obtained as the
momentum distribution of hadrons shows. In addition, nuclei are finite size
systems where the surface plays an important role. This can easily be  seen
inspecting the Weizs\"acker mass formula which gives for infinite matter almost
twice the binding energy/per nucleon as for finite nuclei. 
Therefore complicated 
non-equilibrium transport theories have to be employed and the conclusion 
on the nuclear EoS can only be indirect, in determining the EoS for those
potentials which give best agreement with the heavy ion results.

In order to determine the energy which is necessary to compress
infinite nuclear matter in thermal equilibrium by heavy ion
reactions in which no equilibrium is obtained one chooses the
following strategy: The transport theory calculates the time
evolution of the quantal particles described by  Gaussian wave
functions. The time evolution is given by a variational principle
and the equations one obtains for this choice of the wave function
are identical to the classical Hamilton equations where the
classical two-body potential is replaced by the expectation value
of a Skyrme potential. The Skyrme potential is a simple
approximation to the real part of the Br\"uckner $G$-matrix which
is too complicated for performing simulations of heavy ion collisions.
For this potential the potential energy in infinite nuclear matter
is calculated. To determine the nuclear EoS we average this
(momentum-dependent) two-body potential over the momentum
distribution of a given temperature $T$ and add to it the kinetic
energy. Expressed as a function of the density we obtain the
desired nuclear EoS $E/A(\rho,T)$. The potential which we use 
has five parameters. Four of them  are fixed by the binding
energy per nucleon in infinite nuclear matter at $\rho_0$ and the 
optical potential which has been measured in pA reactions \cite{har}. 
The only parameter which has been not determined by experiments yet is 
the compressibility $\kappa$ at $\rho_0$. For $\kappa < 250$ MeV one 
calls the EoS soft, whereas an EoS is called hard for
$\kappa >$ 350 MeV. Once the parameters are fixed we use the two-body potential with
these parameters in the transport calculation. There is an
infinite number of two-body potentials which give the same
EoS because the range of the potential does not play
a role in infinite matter. The nuclear surface measured in
electron scattering on nuclei fixes the range, however, quite
well. 

The different transport theories give  quite comparable results for 
the bulk part but it is difficult to model the surface. (In these simulations 
there is no surface in the strict sense. Each nucleon contributes 
to the density by its Gaussian wave function and the positions of the
hadrons in the course of the reaction determine the surface 
as well as the density gradients.)

The in-plane flow is caused by the density gradient and
hence the numerical value depends on how good the surface of the nucleus 
can be modeled during the reaction. Already small density fluctuations, 
which are difficult to control, change the value of the in-plane flow 
considerably. Therefore the second approach, the determination
if the EoS by measuring the in-plane flow, has not produced
conclusive results yet\cite{ant}.

The third approach, to measure the EoS by means of the $K^+$ yield, 
depends on bulk properties of matter and surface fluctuations have
no influence. Here the different transport theories have converged. 
This was possible due to a special workshop
at the ECT* in Trento/Italy where the authors of the different simulation
codes have discussed
their approaches in detail and have unified most of the input quantities. The
results of this common effort have been published in \cite{sqm04}.
As an example we display here the $K^+$ $p_t$ spectra
at midrapidity obtained in the different transport theories at different
energies. Because with each $K^+N$ rescattering collision the slope
of the $K^+$ spectra changes the slope of the $p_t$ spectra encodes not only
the $K^+$ momentum distribution at the time point of production but also
the distribution of the number of rescatterings. It is therefore all but
trivial. Without the $KN$ potential the slopes are almost identical and even
the absolute yield, which depends on a correct modeling of the Fermi motion of
the nucleons, is very similar. If we include the $KN$ interaction which is not
identical in the different approaches (see \cite{sqm04}) we still observe a very
similar slope for most of the programs.

\begin{figure}
\vspace*{-.5cm} 
\epsfig{file=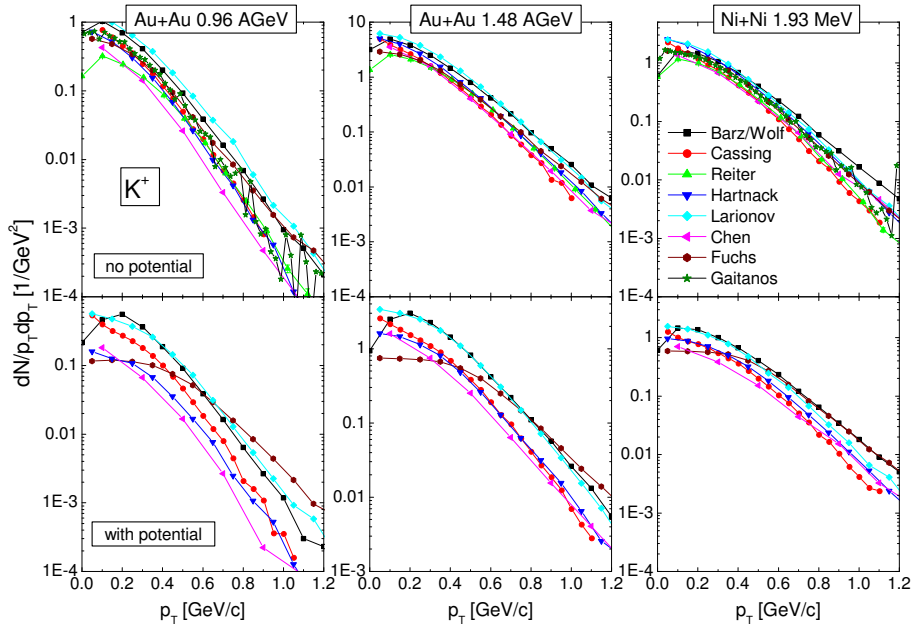,width=12cm}
\caption{Final $K^+$ transverse momentum distribution at $b$=1 fm, $|y_{cm} < 0.5|$ and with an
enforced $\Delta$ lifetime of 1/120 MeV (top row without, bottom row with KN
potential) in the different approaches \cite{sqm04}.}
\label{kp-pt}
\end{figure}

Due to this progress the simulation programs can now be used to extract
up to now  theoretically inaccessible information like the hadronic 
EOS \cite{prl}.
Three independent experimental observables, the ratio of the excitation 
functions of the $K^+$ production for Au+Au and for C+C~\cite{sturm,fuchs}, 
the dependence of the $K^+$ yield on the number of participants and the
excitation function of this dependence can be simultaneously reproduced
if in these transport theories  the nucleons interact with potential which 
yield in infinite matter in equilibrium a compressibility  of 
of the EoS of $\kappa  \approx  200$ MeV.
Large compressibility moduli yield results which disagree with all 
three observables.

This value of $\kappa$ extracted from the $K^+$ production which is
sensitive to nuclear matter around 2.5$\rho_0$ is very similar to that 
extracted by the study of monopole vibrations at $\rho_0$ \cite{you,mon}.

It is not sufficient to determine the compressibility modulus. One has to
demonstrate as well that its numerical value is robust, i.e. 
that the different implementations of yet unsolved physical questions,
like the $N\Delta \rightarrow K^+ \Lambda N$ cross section, the
$KN$ interaction as well as the life time of the nuclear
resonances in the hadronic environment do not affect its value.

\begin{figure}[htb]
\begin{tabular}{cc}
\epsfig{file=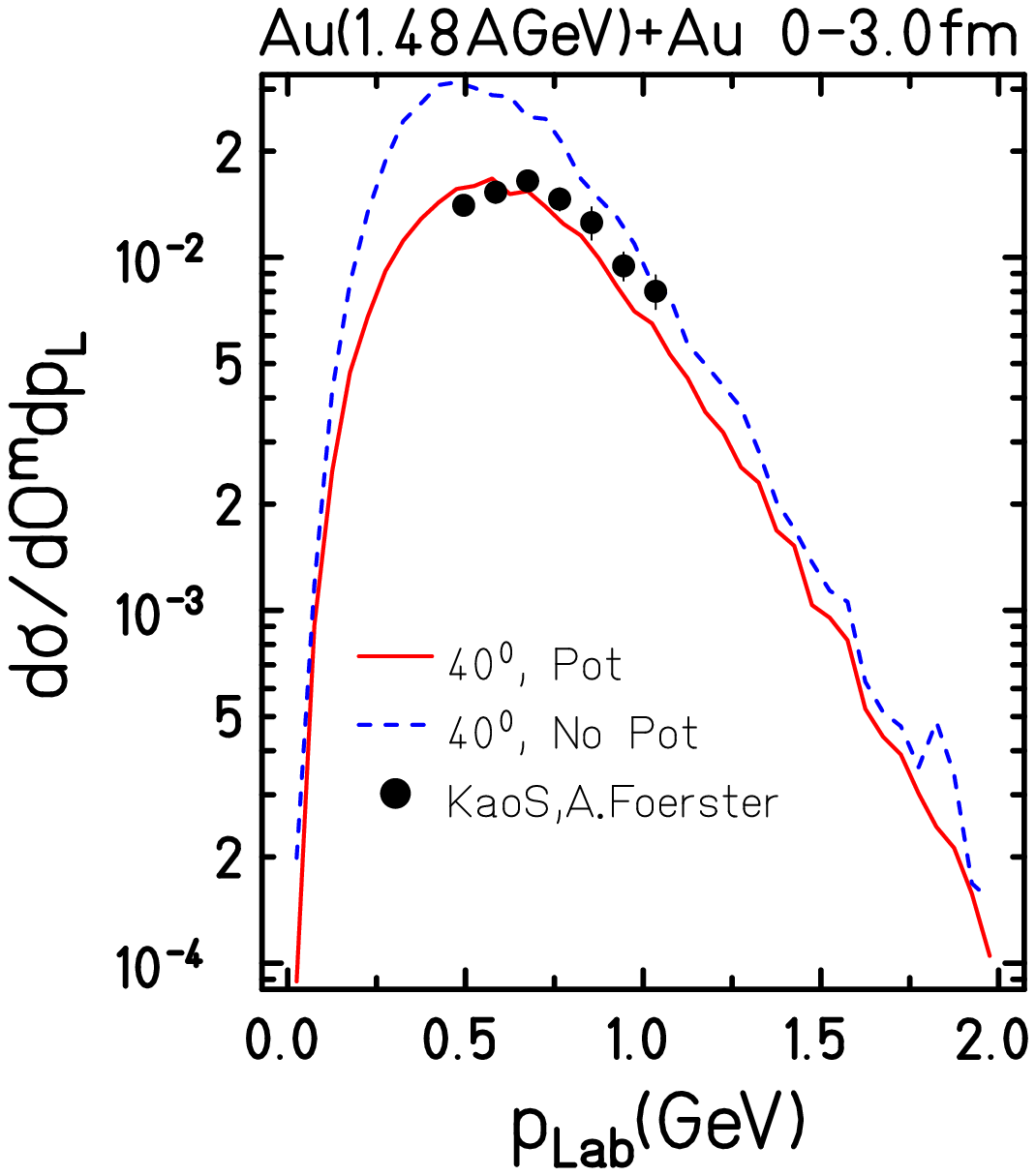,width=0.5\textwidth} &
\epsfig{file=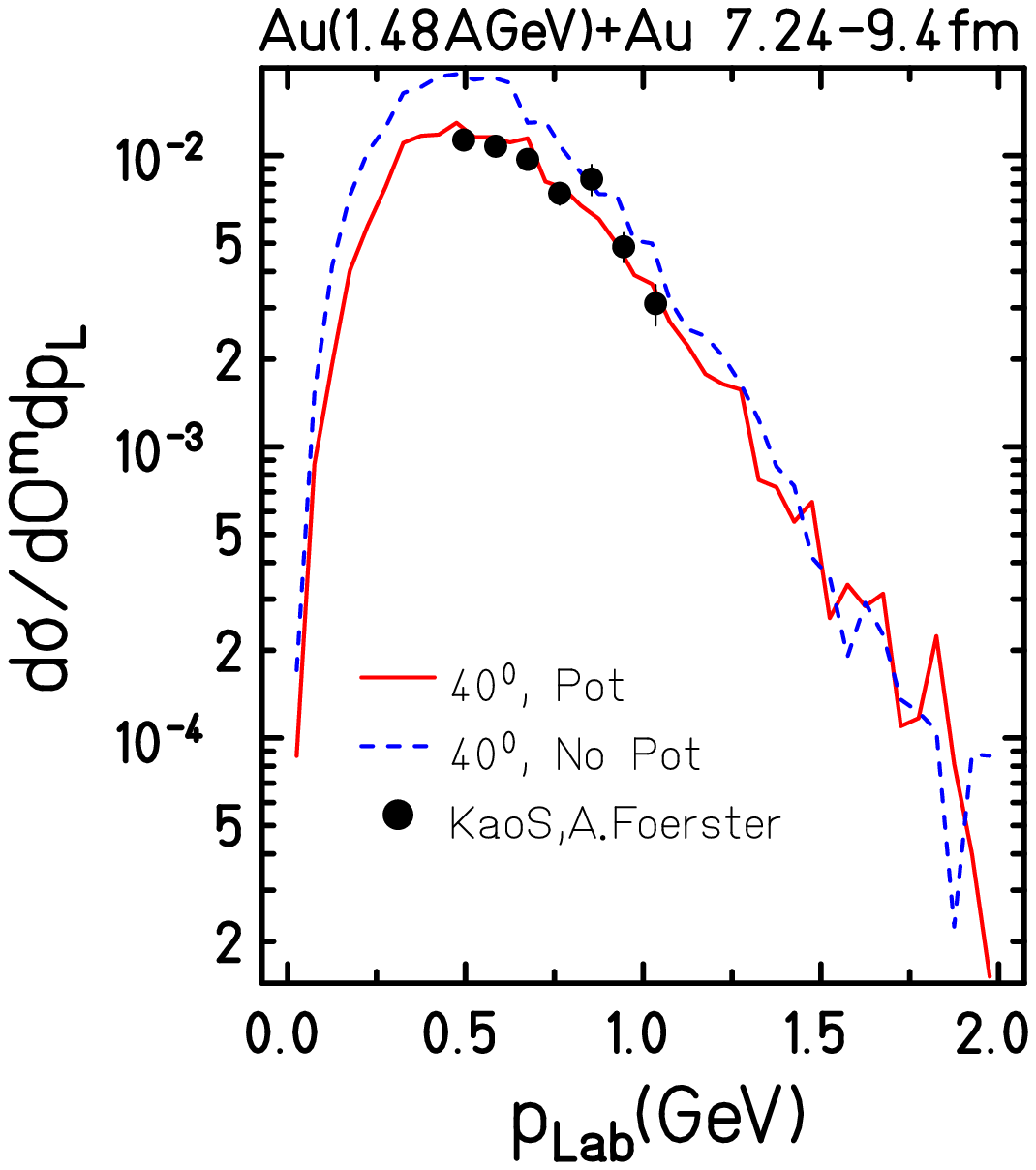,width=0.5\textwidth} \\
\end{tabular}
\caption{$K^+$ spectra for different centrality bins as compared with
(preliminary] experimental data from the KaoS collaboration}
\label{nuc-col}
\end{figure}

We employ the Isospin Quantum Molecular Dynamics (IQMD) with
momentum dependent forces. All details of the standard version of
the program may be found in \cite{har}. In addition we have implemented
for this calculation all cross sections which yield to the production
of $K^+$ as well as the elastic and the charge exchange 
$KN\rightarrow KN$ reactions. The parametrization of the cross section may be
found in \cite{sqm04}. In the standard version the $K^+N$ potential
leads to an increase of the $K^+$ mass in matter,
 $m^K(\rho) = m^K_0(1-0.075 \frac{\rho}{\rho_0})$, in agreement with
recent self-consistent calculations of the spectral function of the
$K^+$ \cite{lu}. The $\Lambda$ potential is 2/3 of the nucleon
potential, assuming that the s quark is inert. The calculations
reproduce the experimental data quite well as can be seen in fig.
\ref{nuc-col} where we compare the experimental and theoretical
$K^+$ spectra for different centrality bins and for 1.48 AGeV Au+Au.
This figure shows as well the influence of the $K^+N$ potential
which modifies not only the overall multiplicity of $K^+$ due to the
increase of the in medium mass but also the spectral form confirming
the complexity of the transverse momentum spectrum.

In order to minimize the experimental systematical errors and the
consequences of theoretical uncertainties the KaoS collaboration has
proposed to study not directly the excitation function of the $K^+$ yield 
but that of the yield ratio of heavy to light systems \cite{sturm}.
Calculations have shown that ratios are much less sensitive to
little known input parameters because these affect both systems in a
rather similar way. We have shown in fig. \ref{nuc-col} that
the absolute yields are well reproduced in our simulations.
Therefore we can use this ratio directly for a quantitative 
comparison with data. The ratio of the 
$K^+$ yields obtained in C+C and Au+Au collisions is quite sensitive 
to the EoS because in Au+Au collisions densities up to 3 $\rho_0$ 
(depending on the EoS) are reached whereas in C+C collisions compression 
is practically absent due to less stopping.

\begin{figure}
\vspace*{-.5cm} 
\epsfig{file=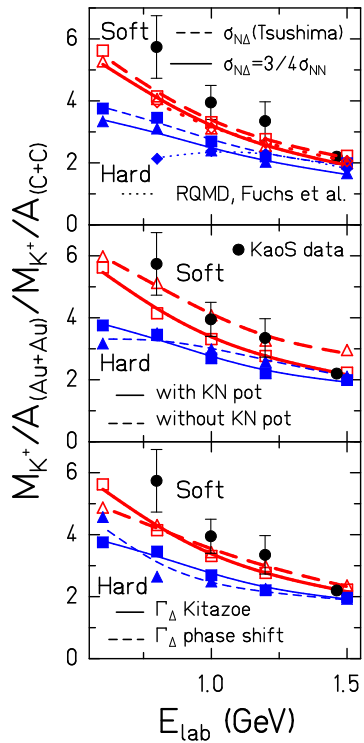,width=7.0cm}
 \caption{Comparison of the measured excitation
function of the ratio of the $K^+$ multiplicities per mass number
$A$ obtained in Au+Au and in C+C reactions (Ref.~\cite{sturm})
with various calculations. The use of a hard EoS is denoted by
thin (blue) lines, a soft EoS by thick (red) lines. The calculated
energies are given by the symbols, the lines are drawn to guide
the eye. On top, two different versions of the $N\Delta
\rightarrow K^+\Lambda N$ cross sections are used. One is based on
isospin arguments \cite{ko}, the other is determined by a
relativistic tree level calculation \cite{tsus}. The calculation
by Fuchs \cite{fuchs} are shown as dotted lines. Middle: IQMD
calculations with and without $KN$ potential are compared. Bottom:
The influence of different options for the life time of  $\Delta$
in matter is demonstrated.} \label{ratio}
\end{figure}

Figure \ref{ratio} shows the comparison of the measured ratio of the
$K^+$ multiplicities obtained in Au+Au and C+C reactions
\cite{sturm} together with transport model calculations as a
function of the beam energy \cite{prl}. We see, first of all in the top row,
that the excitation function of the yield ratio depends on the
potential parameters (hard EoS: $\kappa$ = 380 MeV, thin lines and
solid symbols, soft EoS: $\kappa$ = 200 MeV , thick lines and open
symbols) in a quite sensible way and - even more essential -
that the prediction in the standard version of the simulation
(squares) for a soft and a hard EoS potential differ much more than
the experimental uncertainties. The calculation of Fuchs et
al.~\cite{fuchs} given in the same graph, agrees well with our
findings.

This observation is, as said, not sufficient to determine the
potential parameters uniquely because in these transport theories
several not precisely known processes are encoded. For these
processes either no reliable theoretical prediction has been
advanced or the different approaches yield different results for the
same observable. Therefore, it is necessary to verify that these
uncertainties do not render our conclusion premature. There are 3
identified uncertainties: the $\sigma_{N\Delta \to K^+}$ cross
section, the density dependence of the $K^+N$ potential and the
lifetime of $\Delta$ in matter if produced in a collisions with a
sharp energy of two scattering partners. We discuss now how these
uncertainties influence our results:

Figure \ref{ratio}, top, shows as well the influence of the unknown
$N\Delta \rightarrow K^+ \Lambda N$ cross section on this ratio. We
confront the standard IQMD option (with cross sections for $\Delta
N$ interactions from Tsushima et al.~\cite{sqm04}) with another
option, $\sigma(N\Delta) = 3/4 \sigma(NN)$~\cite{ko}, which is based
on isospin arguments and has been frequently employed. Both cross
sections differ by up to a factor of ten and change significantly
the absolute yield of $K^+$ in heavy ion reactions but do not change
the shape of the ratio.

The middle part demonstrates the influence of the kaon-nucleon
potential which is not precisely known at the densities obtained
in this reaction. The uncertainties due to the $\Delta$ life time
are discussed in the bottom part. Both calculations represent the
two extreme values for this lifetime \cite{sqm04} which is
important because the disintegration of the $\Delta$ resonance
competes with the $K^+$ production.

Thus we see that these uncertainties do not influence the
conclusion that the excitation function of the ratio is quite
different for a soft EoS potential as compared to a hard one and that
the data of the KaoS collaboration are only compatible with the
soft EoS. The only possibility to change this
conclusions is the assumption that the cross sections are
explicitly density dependent in a way that the increasing density
is compensated by a decreasing cross section. 
It would have a strong influence on other observables which are
presently well predicted by the IQMD calculations.

\begin{figure}[htb]
\begin{tabular}{cc}
\epsfig{file=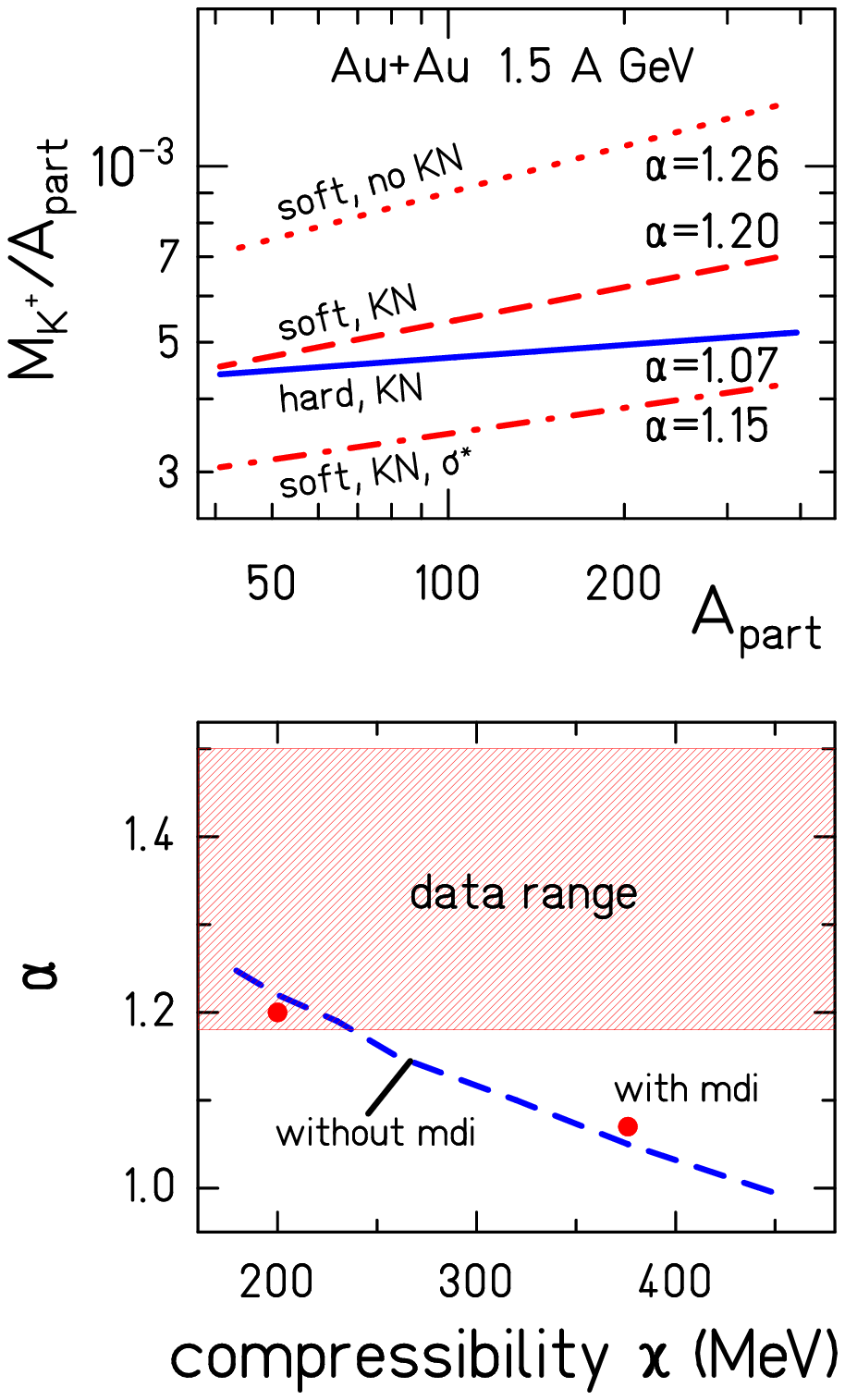,width=5.9cm}
\epsfig{file=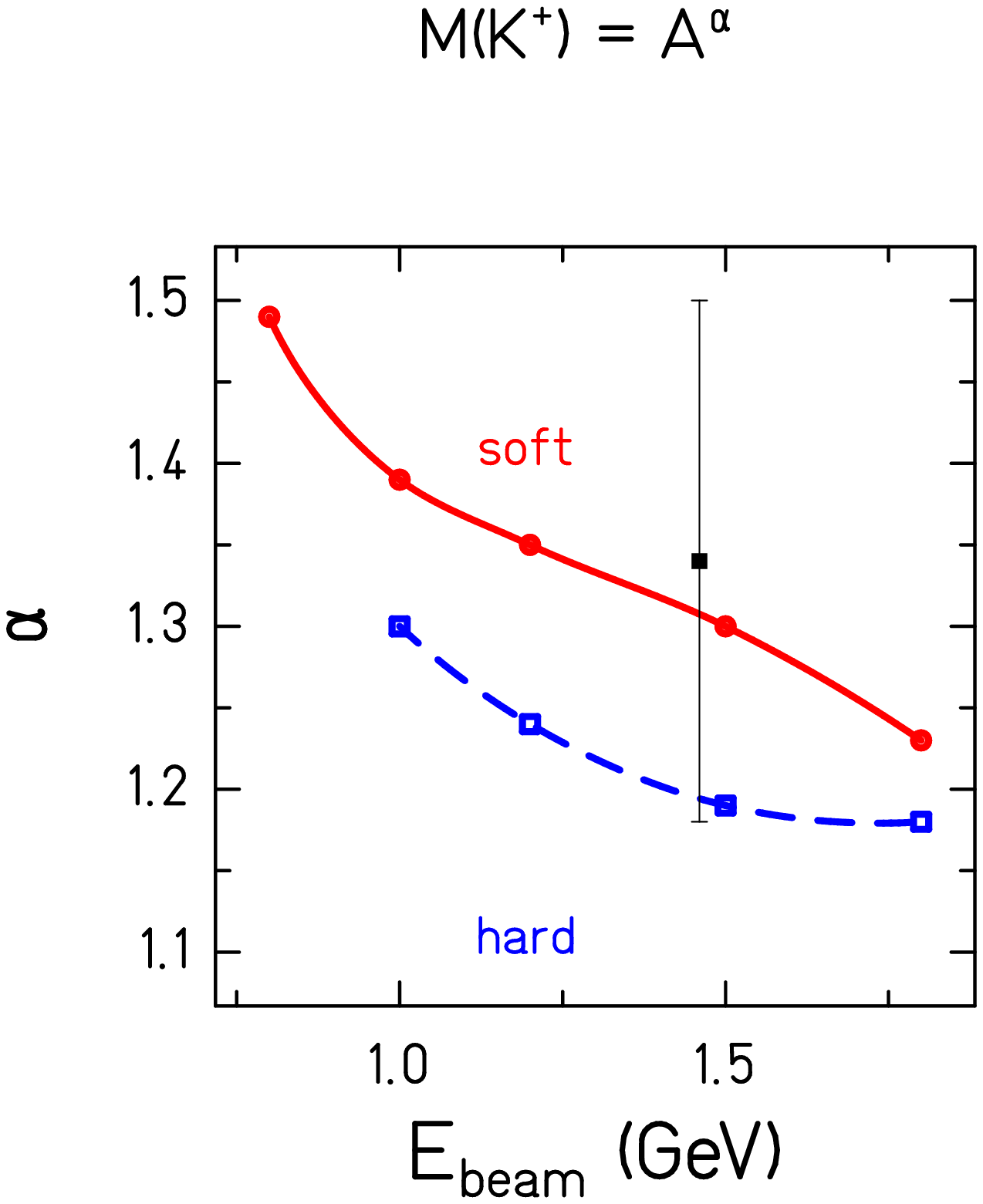,width=5.9cm}
\end{tabular}
\caption{Dependence of the $K^+$ scaling on the nuclear EoS.
We present this dependence in form of $M_{K^+}=A_{{\rm
part}}^\alpha$. On the top the dependence of $M_{K^+}/A_{{\rm
part}}$ as a function of $ A_{{\rm part}}$ is shown for different
options: a ``hard'' EoS with $KN$ potential (solid line), the
other three lines show a ``soft'' EoS, without $KN$ potential and
$\sigma(N\Delta)$ from Tsushima~\cite{tsus} (dotted line), with
$KN$ potential and the same parametrization of the cross section
(dashed line) and with $KN$ potential and $\sigma(N\Delta) = 3/4
\sigma(NN)$. On the bottom the fit exponent $\alpha$ is shown as a
function of the compressibility for calculations with
momentum-dependent interactions (mdi) and for static interactions
(dashed line)\cite{har}. On the right hand side we compare the energy dependence
of the fit exponent $\alpha$ for the two EoS.} \label{part}
\end{figure}

The compression which can be obtained in heavy ion reactions
depends on the impact parameter or, equivalently, on the
experimentally accessible number of participating nucleons. 
Therefore by varying the impact
parameter we can test the EoS at different densities. This dependence 
should be different for different EoS. This is indeed the case for the result
of the simulations as seen in Fig.~\ref{part}, top,  where we display 
the kaon yield $M_{K^+}/A_{{\rm part}}$ for Au+Au collisions
at 1.5 $A$ GeV as a function of the participant number ${A_{{\rm
part}}}$ and for different options: standard version
(soft, $KN$), calculations without kaon-nucleon interaction (soft,
no $KN$) and with the isospin based $N\Delta \rightarrow N\Lambda
K^+$ cross section (soft, $KN$, $\sigma^*$).  A variation of 
the KN potential as well as of the $K^+$ production cross section
change the dependence of the $K^+$ yield on the number of participants,
which can be parametrized by the form $M_{K^+}=A_{{\rm part}}^\alpha$,
only little. On the contrary, if we apply a hard EoS, the slope value
$\alpha$ changes considerable and is outside of the values which are compatible
with the experimental results, as shown in the middle part of the figure.  
In this figure we display as well the insensitivity of our
result to the momentum dependence of the nucleon nucleon interaction.
As long as the compressibility is not changed the results of our calculations
are very similar independent on whether we have a static or 
a momentum dependent NN potential. Hence the dependence of the $K^+$ yield
on the number  of participants is also a robust variable for the
determination of the EoS which supports our earlier conclusion that the EoS 
is soft. 
Another confirmation that only a soft EoS describes the experimental data
is the beam energy dependence of the fitted exponent $\alpha$ which 
is displayed in the right part of fig. \ref{part}. The  data, 
which follow the curve for a soft equation of state, will soon be published 
\cite{kao}.

In conclusion, we have shown that earthbound experiments have now reached a
precision which allows to determine the hadronic EoS. The experimental results 
for the three observables which are most sensitive to the hadronic EOS
are only compatible with theory if the hadronic EoS is soft.
This conclusion is robust. Little known input quantities do not influence 
this conclusion. The observation of a neutron star with twice the solar
mass seems to contradict this conclusion. It points toward a hard
hadronic EoS. Both results are quite new and one has not to forget that
we are comparing non equilibrium heavy ion reactions where about the 
same number of protons and neutrons are present and where mesons and baryon
resonances are produced with cold neutron matter in equilibrium.
In addition this contradiction depends also on the prediction that
the observed star mass excludes the formation of quark matter in the 
interior, a consequence of the suggested EoS of quark matter which is
still rather speculative.
  
To solve this contradiction is certainly a big challenge for both communities in the
near future.


\begin{thebibliography}{99}
\bibitem{merger} Sasa Ratkovic, Madappa Prakash, James M.Lattimer
Submitted to ApJ, astro-ph/0512136
\bibitem{pod}Ph. Podsiadlowski, J. D. M. Dewi, P. Lesaffre, J. C. Miller,
W. G. Newton, J. R. Stone
Mon.Not.Roy.Astron.Soc. 361 (2005) 1243-1249,
astro-ph/0506566
\bibitem{lat}J.M. Lattimer, M. Prakash
Science Vol. 304 2004 (536-542)
\bibitem{lat1}J.M. Lattimer and M. Prakash, ApJ {\bf 550} (2001) 426;
A.W. Steiner, M. Prakash and J.M. Lattimer,
Phys. Lett. {\bf B486} (2000) 239;
M. Alford and S. Reddy,
Phys. Rev. D. {\bf 67} (2003) 074024.
\bibitem{jan} H.-Th. Janka, R. Buras, F.S. Kitaura Joyanes, A. Marek,
M. Rampp \\
Procs. 12th Workshop on Nuclear Astrophysics, Ringberg Castle,
March 22-27, 2004, astro-ph/0405289
\bibitem{weber}Fridolin Weber,  Prog.Part.Nucl.Phys. 54 (2005) 193-288
\bibitem{star} David J. Nice, Eric M. Splaver, Ingrid H. Stairs, Oliver Loehmer,
 Axel Jessner, Michael Kramer, James M. Cordes, submitted to ApJ,
 astro-ph/0508050
\bibitem{mai}C. Maieron, M. Baldo, G.F. Burgio, H.-J. Schulze
Phys.Rev. D70 (2004) 043010
\bibitem{bal} M. Baldo, M. Buballa, G.F. Burgio, F. Neumann, M. Oertel, H.-J. Schulze
Phys.Lett. B562 (2003) 153-160
\bibitem{gas}F. Gastineau, R. Nebauer, J.Aichelin, Phys.Rev. C65 (2002) 045204
\bibitem{you}D.H. Youngblood, H.L. Clark and Y.-W. Lui, Phys. Rev. Lett {\bf
84} (1999) 691.
\bibitem{mon}  J. Piekarewicz, Phys. Rev. {\bf C69} (2004) 041301
and references therein.
\bibitem{sto} H. St\"ocker and W. Greiner,
\newblock  Phys.~Reports~{\bf 137}, 278 (1986) and references therein.
\bibitem{aik} J. Aichelin and C.M. Ko, Phys. Rev. Lett. {\bf 55} (1985) 2661.
\bibitem{prl}Ch. Hartnack, H. Oeschler, J. Aichelin, Phys.Rev.Lett. 96 (2006) 
012302
\bibitem{har} C. Hartnack et al., Eur. Phys. J. {\bf A1} (1998) 151.
\bibitem{ant} A. Andronic et al., Phys. Lett. {\bf B612} (2005) 173.
\bibitem{sqm04} E.E. Kolomeitsev et al., J.Phys. G31 (2005) S741
\bibitem{sturm} C. Sturm et al., (KaoS Collaboration), Phys. Rev. Lett. {\bf 86}
(2001) 39.
\bibitem{fuchs}
C. Fuchs et al., Phys. Rev. Lett {\bf 86} (2001) 1794.
\bibitem{lu} C.L. Korpa and M.F.M. Lutz, submitted to Heavy Ion Physics,
nucl-th/0404088.
\bibitem{habil} C. Hartnack and J. Aichelin, Proc. Int. Workshop XXVIII
on Gross prop. of Nucl. and Nucl. Excit., Hirschegg, January 2000
edt. by M. Buballa, W. N\"orenberg, B. Sch\"afer and J. Wambach;
and  to be published in Phys. Rep.
\bibitem{ko}
J.~Randrup and C.M.~Ko,
\newblock Nucl.~Phys. {\bf A 343}, 519 (1980).
\bibitem{tsus} K. Tsushima et al., Phys. Lett. {\bf B337} (1994) 245;
Phys. Rev. {\bf C 59} (1999) 369. 
\bibitem{foer} A. F\"orster et al., (KaoS  Collaboration), Phys. Rev. Lett. {\bf
31} (2003) 152301; J. Phys. G. {\bf 30} (2004) 393; A.~F\"orster,
Ph.D.~thesis, Darmstadt University of Technology, 2003.
\bibitem{kao} KaoS collaboration, private communication and to be published.

\end{thebibliography}
\end{document}